\documentclass{ws-ijmpa}
\newcommand{\be}{\begin{equation}}
\newcommand{\ee}{\end{equation}}
\newcommand{\ba}{\begin{array}}
\newcommand{\ea}{\end{array}}
\newcommand{\bea}{\begin{eqnarray}}
\newcommand{\eea}{\end{eqnarray}}
\newcommand{\pro}{\partial}

\newcommand{\hD}{{\hat D}}

\newcommand{\hR}{{\hat R}}

\newcommand{\difrac}{\displaystyle\frac}
\newcommand{\nn}{\nonumber}

\newcommand{\astK}{\stackrel{\ast}{K}}
\newcommand{\TS}{\stackrel{\top}{S}}
\newcommand{\TTS}{\stackrel{\top\!\!\top}{S}}
\newcommand{\TR}{\stackrel{\top}{R}}
\newcommand{\TTR}{\stackrel{\top\!\!\top}{R}}
\begin{document}

\markboth{Y.M. Cho, D.G. Pak, B.S. Park}
{A minimal model of Lorentz gauge gravity with dynamical torsion}

%
\catchline{}{}{}{}{}
%

\title{A MINIMAL MODEL OF LORENTZ GAUGE GRAVITY WITH DYNAMICAL TORSION
}

\author{Y. M. Cho
}

\address{School of Electrical and Computer Engineering\\
    Ulsan National Institute of Science and Technology,
    Ulsan 689-798, Korea\\
ymcho@unist.ac.kr}

\author{D. G. Pak \footnote{Institute of Applied Physics, Uzbekistan National University,
        Vuzgorodok, Tashkent, 100-174, Uzbekistan}}

\address{Center for Theoretical Physics, Seoul National
University\\
Seoul, 151-742, Korea\\
dmipak@gmail.com}

\author{B. S. Park}

\address{College of Natural Sciences, Seoul National
University\\
Seoul, 151-742, Korea\\
parkbs2@snu.ac.kr}

\maketitle

\begin{history}
\received{Day Month Year}
\revised{Day Month Year}
\end{history}

\begin{abstract}
A new Lorentz gauge gravity model with $R^2$-type Lagrangian is proposed.
In the absence of classical torsion the model admits a topological phase
with an arbitrary metric. We analyze the equations of motion in constant
curvature space-time background using the Lagrange formalism
and demonstrate that the model possesses a minimal set of dynamic degrees of freedom
for the torsion. Surprisingly, the number of torsion dynamic degrees of freedom
equals the number of physical degrees of freedom for the metric tensor.
An interesting feature of the model is that the spin two mode of torsion
becomes dynamical essentially due to the non-linear structure of the theory.
We perform covariant one-loop quantization of the model for a special case
of constant curvature space-time background. We treat the contortion
as a quantum field variable whereas the metric tensor is kept as a classical object.
We discuss a possible mechanism of an emergent Einstein gravity as a part of
the effective theory induced due to quantum dynamics of torsion.

\end{abstract}

\ccode{PACS numbers: 04.60.-m, 11.30.Cp}

\section{Introduction}

The gauge approach to gravity based on
gauging Lorentz and Poincare groups\cite{uti}\cdash\cite{cho12}
was developed as a powerful tool in attempts to construct
a consistent quantum theory of gravity in the framework of field
theory formalism\cite{anttomb}\cdash\cite{martell}.
The extension of gauge gravity models to the case
of Riemann-Cartan space-time geometry reveals wide possibilities
towards formulation of numerous models of quantum gravity
with torsion\cite{ivan1}\cdash\cite{shapiro}.
Recently, a Lorentz gauge model of gravity with Yang-Mills type Lagrangian
including contortion (torsion) has been developed further\cite{pak}.
It has been proposed that the Einstein gravity
with a cosmological term can be induced as an effective theory
due to quantum corrections of contortion.
In that model the space-time metric is treated as a fixed classical
field while the contortion supposed to be a quantum field.
Such a treatment of the metric is not satisfactory from the
conceptual point of view since one has to assume the existence
of a classical space-time with a metric given a priori. In other words,
we encounter the problem of space-time background dependence which is similar
to the space-time dependence problem in superstrings where the string
lives in a pre-defined fiducial space-time.
One possible way to resolve this problem is to generalize the
Lorentz gauge model by extending the gauge group to Poincare one.
In that case the gauge potential of the Poincare group, the vielbein,
becomes dynamical on equal footing with torsion.
Another interesting possibility is to look for
a quantum gravity model which has only one basic quantum field, contortion,
whereas the metric describes a topological phase in the absence of
contortion at classical level.
In the topological phase the
torsion can be unobservable quantity at classical level
(this assumption may have physical sense if there is some
mechanism of hiding the classical torsion like the Meissner effect\cite{hanson})
but manifests itself through the quantum dynamics
which triggers the phase transition to the phase
of an effective theory where the metric obtains its
dynamic content.

   In the present paper we propose a special $R^2$-type model of Lorentz gauge
gravity which admits a topological phase at classical level
and has non-trivial quantum dynamics of torsion.
The proposed model is minimal in a sense that only the
contortion possesses dynamic degrees of freedom
whereas the metric does not.
We will demonstrate that the contortion
has six propagating modes with spins $J=(2;1;0;0)$, exactly the same number of
physical degrees of freedom the metric tensor has in general.

Notice, the structure of Poincare gauge theory with a most general $R+R^2$ type Lagrangian
was studied in Refs. \refcite{hayashiI}--\refcite{kuhfuss}. A special case with vanishing
linear in curvature term was considered partially in Ref. \refcite{nieuw}.
What we consider in the present paper represents
a strongly degenerate case of the theory, and its
analysis was missed in previous studies. Moreover,
in our analysis we take into account the effects
conditioned by the curved space-time. We show that
these effects are important and lead to appearance
of the propagating torsion mode with spin two. In a fact,
introducing the non-flat space-time background
we estimate indirectly the effect of the non-linear structure
of the full theory. So that, we expect that our model
even with flat metric has very non-trivial dynamic content for the
contortion due to non-linearity of the equations of motion.
The non-linearities of constraints in Poincare gauge theory
are very important and can change the number of constraints
and degrees of freedom\cite{nesterAPP}\cdash\cite{nesterIJMP02}

In section II, we consider a general Lagrangian
quadratic in Riemann-Cartan curvature
which admits the topological phase
in the limit of vanishing classical torsion.
In section III, we study the dynamic content
of the model in the framework of Lagrange formalism.
By explicit solving all constraints among the equations of motion
we show that the contortion possesses propagating modes.
A covariant quantization scheme and the analysis of the structure of the
one-loop effective Lagrangian is presented in Section IV.
The last section contains discussion on possibility of
inducing the Einstein-Hilbert term and cosmological constant by quantum
torsion corrections.

\section{Lorentz gauge gravity model with a topological phase}

Let us start first with the main outlines of Riemann-Cartan geometry.
The basic geometric objects in Poincare gauge models of
gravity\cite{uti}\cdash\cite{sciama,cho11,cho12} are
the vielbein $e_a^i$ and the general Lorentz affine connection $A_i^{~cd}$.
The infinitesimal Lorentz transformation of the
vielbein $e_a^i$ is given by
\bea
&&\delta e_a^i= [{\bf \Lambda},e_a^i]=\Lambda_a^{~b} e_b^i,
\eea
where ${\bf \Lambda} \equiv \Lambda_{cd} \Omega^{cd}$ is a Lie algebra
valued gauge parameter, and $\Omega^{cd}$ is a generator of the Lorentz
Lie algebra. We use ($i,j,k,...=0,1,2,3$) to denote coordinate indices,
and ($a,b,c,...=0,1,2,3$) for Lorentz frame indices. We assume that the vielbein is
invertible and the signature of the flat metric $\eta_{ab}$
is Minkowskian, $\eta_{ab}={\rm diag} (+---)$.

The covariant derivative with respect to the Lorentz group
transformation is defined in a standard manner
\bea
D_a=e_a^i (\pro_i + g{\bf A}_i) ,
\eea
where ${\bf A}_i\equiv A_{i cd} \Omega^{cd}$ is a general
affine connection taking values in the Lorentz Lie algebra, and
$g$ is a new gravitational gauge coupling constant.
For brevity of notation we will use a
redefined connection which absorbs the coupling constant. The
original Lorentz gauge transformation of the connection ${\bf A}_i$ has the form
\bea
 \delta {\bf A}_i&=&-\pro_i {\bf \Lambda} - [{\bf A}_i, {\bf \Lambda}].
\label{eq:deltaA}
\eea

The affine connection $A_{i cd}$ can be rewritten
as a sum of the Levi-Civita spin connection $\varphi_{i c}^{~~d}(e) $
and the contortion $K_{i c}^{~~d}$
\bea
A_{i c}^{~~d} &=& \varphi_{i c}^{~~d}(e) + K_{i c}^{~~d}, \label{split} \\
\varphi_{i a}^{~~b}(e)&=&-\difrac{1}{2} (e^{j b} \pro_i e_{j a}-
e_a^j e_i^c \pro^b e_{j c}+\pro_a e_i^b
-e_a^j \pro_i e_j^b+e^{j b} e_i^c \pro_a e_{j c}-\pro^b e_{i a}). \nn
\eea

The torsion and curvature tensors are defined in a standard way
\bea
&& [D_a,D_b]=-T_{ab}^{\,\,c} D_c-{\bf R}_{ab} , \nn \\
&& T_{ab}^{\,\,c}=K_{ba}^{~c}-K_{ab}^{~c},
\eea
where, ${\bf R}_{ab} \equiv R_{ab cd} \Omega^{cd}$.
Under the decomposition (\ref{split})
the Riemann-Cartan curvature is splitted into two parts
\bea
&& R_{abcd}=\hat R_{abcd}+\tilde R_{abcd}, \nn \\
&& \hat R_{abcd}=\hat D_{\underline b} \varphi_{\underline a \underline c
\underline d}+\varphi_{bc}^{~~\,\,e}\varphi_{aed}-(a\leftrightarrow b), \nn \\
&& \tilde R_{abcd}=\hat D_{\underline b} K_{\underline a \underline c\underline d}
+K_{bc}^{~~e} K_{aed}-(a\leftrightarrow b) ,
\eea
where, $\hat D_a$ is a restricted covariant derivative containing only the Levi-Civita connection,
and the underlined indices stand for indices over which the
covariantization is performed.

We exclude from our consideration any quadratic invariants
with a dual Riemann-Cartan curvature which breaks the parity.
The most general quadratic in Riemann-Cartan curvature Lagrangian reads
\bea
{\cal L}=c_1 R_{ab cd} R^{abcd}+c_2R_{ab cd} R^{cdab}+c_3 R_{ab} R^{ab}
+c_4 R_{ab} R^{ba} + c_5 R^2 + c_6 A_{abcd}^2, \label{Lgeneral}
\eea
where the last term is an additional invariant
which appears in Riemann-Cartan space-time.
The tensor $A_{abcd}$ is defined as follows\cite{hayashiI}
\bea
A_{abcd}=\dfrac{1}{6} (R_{abcd}+R_{acdb}+R_{adbc}+R_{bcad}+R_{bdca}+R_{cdab})
\eea
In Riemannian space-time the tensor $A_{abcd}$ vanishes due to Jacobi cyclic identity
\bea
R_{abcd}+R_{acdb}+R_{adbc}\equiv 0 .
\eea

In Riemann-Cartan geometry the proper generalization of the topological Gauss-Bonnet invariant
is given by the Bach-Lanczos density\cite{bach-lanczos,lanczos}
\bea
I_{BL}=R_{ab cd} R^{cdab}-4 R_{ab} R^{ba} + R^2.
\eea
 The properties of the Bach-Lanchos invariant
are described in a detail in Ref. \refcite{hayashiV}.

We are interested in such a Lagrangian in Riemann-Cartan space-time which is reduced to the
topological Gauss-Bonnet density in the limit of Riemannian geometry.
A proper Lagrangian can be derived from the general expression
(\ref{Lgeneral}) by fitting the parameters $c_i$
as follows
\bea
{\cal L}&=&-\dfrac{1}{4}\big (\alpha R_{abcd}^2+(1-\alpha) R_{abcd}R^{cdab} -4 \beta R_{bd}^2- \nn \\
     &&4(1-\beta) R_{bd}R^{db} +R^2+6 \gamma A_{abcd}^2\big ) ,
\eea
where the parameters $\alpha,\beta, \gamma$ define the
remaining arbitrariness. One can check that the
Lagrangian reduces to the Gauss-Bonnet density
in the limit of Riemannian space-time geometry.
One can rewrite the Lagrangian in a more simple form
\bea
{\cal L}&=&-\dfrac{1}{4}\Big[(\alpha+\gamma)R_{abcd}^2-(\alpha-\gamma)R_{abcd}R^{cdab}
 -4 \beta( R_{bd}^2-R_{bd}R^{db}) + \nn \\
 && 4 \gamma R_{abcd}R^{acdb} +I_{BL} \Big]\label{lastL}.
 \eea
We will demonstrate that the model described by the Lagrangian (\ref{lastL})
admits dynamic degrees of freedom for the contortion only for
special values of the parameters $\alpha,\beta, \gamma$.

\section{Equations of motion, constraints}

A detailed analysis of the equations of motion of
the general $R+T^2+R^2$ type Poincare gauge gravity
for most of non-degenerate cases
was performed in Refs. \refcite{hayashiI,nieuw}.
A Hamiltonian formalism of the Poincare gauge gravity
was considered in Refs. \refcite{blagoj,nikolic}.
For our model, which is strongly degenerate and
has a non-trivial space-time background,
it is much more easier to use the Lagrange formalism instead of the canonical Hamiltonian one\cite{gitman}.
We start with arbitrary parameters $\alpha, \beta, \gamma$ in the initial Lagrangian (\ref{lastL}), and then
we will find constraints on the parameters under which
the contortion obtains propagating modes.
Let us consider linearized equations of motion corresponding to the Lagrangian (\ref{lastL})
\bea
\dfrac{\delta {\cal L}}{\delta K_{bcd}}&&\equiv\dfrac{1}{2}(\alpha+\gamma)(\hD^a \hD_a K_{bcd}-\hD^a \hD_b K_{acd})
     -(\alpha-\gamma)\hD^a \hD_c K_{dab}+ \nn \\
 \gamma&&(\hD^a \hD_a K_{cdb}-\hD^a \hD_c K_{adb}-\hD^a \hD_b K_{cda}-\hD^a \hD_c K_{bda})+\nn \\
\beta&& [\hD_c \hD^a K_{dab}- \hD_c \hD^a K_{bad}+\hD_c \hD_b K_d -\hD_c \hD_d K_b+ \nn \\
 \eta_{bc}&& (\hD^a \hD^e K_{aed} - \hD^a \hD^e K_{dea}-\hD^a \hD_a K_d + \hD^a \hD_d K_a)]
-(c\leftrightarrow d)=0,
\eea
where, $K_d \equiv K^c_{~cd}$.
It is convenient to impose gauge fixing conditions which
are compatible with equations of motion. To fix the Lorentz gauge
symmetry we choose the following constraints
(indices denoted by Greek letters refer to spatial coordinates)
\bea
&& \pro^\beta (K_{\beta 0\delta}-K_{\delta 0\beta})=0, \nn \\
&&(\alpha+\gamma) \pro^\beta K_{\beta\gamma\delta}=\gamma (\pro^\beta K_{\gamma \beta\delta}
-\pro^\beta K_{\delta \beta\gamma}),  \nn \\
&& \pro^\beta \pro^\delta K_{\beta 0 \delta}=0. \label{G3}
\eea
The consistence of these gauge conditions with the Lagrange equations
will be verified during solving all equations of motion.

The equation of motion $\delta {\cal L}/\delta K_{00\delta}=0$ with
a flat metric and non-vanishing parameter $\beta$
represents a most stringent constraint which
implies that the transverse part of the vector field $K_{\nu\nu\delta}$
vanishes identically. Since we are interested in finding propagating modes for the
spin one vector field even in the flat space-time
we will consider in the following only the case $\beta=0$.
It should be stressed, that the presence of curved space-time
and non-linearity of the theory is an essential factor which can not
be ignored. The perturbative analysis of linearized equations of motion
in the flat space-time may lead to incorrect results in degenerate cases.
For that reason we will study the equations of motion
in the curved space-time, with a non-flat metric,
while keeping a linearized approximation in contortion field.
Such a treatment allows us to describe the
main features of our model since the non-trivial
Levi-Civita connection $\varphi_{i a}^{~~b}(e)$
can be treated as a background part of the general Lorentz connection.
For simplicity we choose the background space-time
as a Riemannian space-time of covariant constant curvature
\bea
\hat R_{abcd}= \difrac{1}{12} \hat R (\eta_{ac} \eta_{bd}-\eta_{ad}\eta_{bc}).
\label{background}
\eea
In constant curvature background space-time
one can rewrite the equations of motion in normal coordinates
keeping only linear terms in Ricci scalar $\hat R$.
The metric tensor, vielbein and Levi-Civita connection have the following
decomposition in normal coordinates
\bea
g_{mn}&=&\eta_{mn}+\dfrac{2\rho}{3} (\eta_{mn} x^k x_k-x_m x_n), \nn \\
e_m^a&=& \delta^a_m +\dfrac{\rho}{3} (\delta_m^a x^k x_k-x^a x_m), \nn \\
\varphi_{ma}^{~~~b}&=& \rho (\eta_{ma} x^b -\delta_m^b x_a),
\eea
where $\rho=\dfrac{\hR}{24}$. Using these relations one can
write down the equations of motion as follows
\bea
&&\dfrac{1}{2}(\alpha+\gamma)(\pro^a \pro_a K_{bcd}-\pro^a \pro_b K_{acd})
     -(\alpha-\gamma)\pro^a \pro_c K_{dab} \nn \\
&& +\gamma(\pro^a \pro_a K_{cdb}-\pro^a \pro_c K_{adb}-\pro^a \pro_b K_{cda}+
               \pro^a \pro_c K_{bda}) \nn \\
&&+\gamma \rho K_{bcd}+(3\alpha-2\gamma) \rho K_{cbd}
-\alpha \rho \eta_{bc} K_d-(c\leftrightarrow d) =0.
\eea
Using the gauge conditions (\ref{G3})
one can rewrite the equations of motion in component form
\bea
\dfrac{\delta {\cal L}}{\delta K_{00\delta}}&\equiv&\Delta K_{00\delta}-\pro^\beta\pro_\delta K_{00\beta}-
   \pro^\beta \pro_0 K_{\beta 0\delta}+\pro^\beta \pro_0 K_{\delta 0 \beta}+ \nn \\
 &&2 \rho K_{00\delta}-\rho K^\nu_{~\nu\delta}=0, \label{E1}
   \eea
   \bea
\dfrac{\delta {\cal L}}{\delta K_{0\gamma\delta}}&\equiv&(\alpha+\gamma)\Delta K_{0\gamma\delta}
   +\alpha(\pro^\beta\pro_\gamma K_{\delta 0\beta}-\pro^\beta\pro_\delta K_{\gamma 0\beta})
   -(\alpha+\gamma)\pro_0\pro^\beta K_{\beta \gamma\delta} +\nn \\
   && \gamma (\pro_\gamma\pro^\beta K_{0\delta \beta}
   -\Delta K_{\gamma 0\delta}-\pro_0\pro^\beta K_{\gamma\delta \beta}-(\gamma\leftrightarrow\delta))
+2 \gamma \rho K_{0\gamma\delta}+\nn \\
&&(3 \alpha-2\gamma)\rho(K_{\gamma 0\delta}-K_{\delta 0 \gamma}), \label{E2}
    \eea
    \bea
\dfrac{\delta {\cal L}}{\delta K^\nu_{~ 0\nu}}&\equiv& \Delta K^\nu_{~ 0\nu}
         +\pro_0 \pro^\beta K^\nu_{~\nu \beta}=0,\label{E3}
\eea
\bea
\dfrac{\delta {\cal L}}{\delta K^\nu_{~\nu\delta}}&\equiv&\Box K^\nu_{~\nu\delta}-\pro^\nu\pro^\beta K_{\beta\nu\delta}
       -\pro_\delta \pro^\beta K^\nu_{~\nu \beta}-\pro_0\pro^\beta K_{\delta 0 \beta}-\pro_0\pro^\beta K_{0\beta\delta } \nn \\
        &-&\pro_0\pro_\delta K_0+\rho K^\nu_{~\nu\delta}-2 \rho K_{00\delta}=0, \label{E4}
        \eea
        \bea
\dfrac{\delta {\cal L}}{\delta K_{\beta 0\delta}}&\equiv&\alpha[\Delta K_{\beta 0\delta}-\pro_\beta\pro^\alpha K_{\alpha 0\delta}+
 \pro_0\pro_0(K_{\beta 0 \delta}-K_{\delta 0 \beta}) +\pro^\alpha\pro_\delta K_{0\alpha\beta} \nn \\
       &-&\pro_0\pro^\alpha K_{\delta \alpha \beta}-\pro_0\pro_\beta K_{00\delta}+\pro_0\pro_\delta K_{00\beta}
           +3\rho(K_{0\beta\delta}+K_{\delta 0 \beta})+\rho \eta_{\beta\delta}K_0] \nn \\
           &+&\gamma[\Delta (K_{\beta 0\delta}-K_{\delta 0\beta})+
           \pro_0\pro^\alpha(K_{\delta \alpha \beta}-K_{\alpha\delta\beta}) \nn \\
           &+&\pro_\delta\pro^\alpha K_{0\beta \alpha}-\pro_\beta\pro^\alpha K_{0\delta \alpha}+\Delta K_{0\delta\beta}
            +\pro_0\pro^\alpha K_{\beta\delta \alpha}]-\nn \\
            &&2 \gamma \rho K_{0\beta\delta}+2\gamma\rho (K_{\beta 0\delta}
                                                       -K_{\delta 0\beta})=0, \label{E5}
\eea
\bea
\dfrac{\delta {\cal L}}{\delta K_{\beta\gamma\delta}}&\equiv& \alpha[\Box K_{\beta\gamma\delta}
  -\pro_\beta \pro^\alpha K_{\alpha\gamma \delta}-\pro_0\pro_\beta K_{0\gamma\delta}-\pro_0\pro_\gamma K_{\delta 0\beta}
   +\pro_0\pro_\delta K_{\gamma 0\beta} \nn \\
    &-&\pro_\gamma \pro^\alpha K_{\delta \alpha\beta}+\pro_\delta \pro^\alpha K_{\gamma \alpha\beta}
     +3\rho(K_{\gamma \beta\delta}-K_{\delta\beta\gamma})-\rho(\eta_{\beta\gamma} K_\delta-\eta_{\beta\delta}K_{\gamma})]\nn \\
     &+&\gamma [(\Box+2\rho)K_{(\beta\gamma\delta)}+\pro_\gamma\pro^\alpha K_{(\alpha\beta\delta)}-
                        \pro_\delta\pro^\alpha K_{(\alpha\beta\gamma)}-\pro_\beta\pro^\alpha K_{(\alpha\gamma\delta)} \nn \\
    &+&\pro_0\pro_{(\gamma} K_{0\beta\delta)}+\pro_0\pro_{(\gamma} (K_{\delta 0\beta)}-K_{\beta 0\delta)})]=0,\label{E6}
\eea
where, $\Box=\pro^a\pro_a, ~ \Delta=\pro^\alpha \pro_\alpha$ ,
and a cyclic combination over indices enclosed in brackets is assumed,
for instance,
\bea
K_{(\beta\gamma\delta)}=K_{\beta\gamma\delta}+K_{\gamma\delta\beta}+K_{\delta\beta\gamma}.
\eea

We will use the following decomposition of the contortion
$K_{\mu cd}$ into irreducible parts\cite{deser}
\bea
&&K_{\mu\gamma\delta}=\epsilon_{\gamma \delta}^{~~~\rho}\stackrel{\ast}{K}_{\mu\rho}, \nn \\
&&\astK_{\mu\rho}=\TTS_{\mu\rho}+\dfrac{1}{2}(\delta_{\mu\rho} \Delta-\pro_\mu\pro_\rho)\TS
 +(\pro_\mu S_\rho+\pro_\rho S_\mu) +\epsilon_{\mu\rho}^{~~~\sigma} A_\sigma , \nn \\
 &&K_{\mu 0 \rho}=\TTR_{\mu\rho}+\dfrac{1}{2}(\delta_{\mu\rho} \Delta-\pro_\mu\pro_\rho)\TR
 +(\pro_\mu R_\rho+\pro_\rho R_\mu) +\epsilon_{\mu\rho}^{~~~\sigma} Q_\sigma,
\eea
where the superscript $"T"$ stands for traceless components and the superscript $"TT"$
denotes irreducible parts which are traceless and transverse.

The equations of motion (\ref{E1}-\ref{E6})
and gauge fixing conditions (\ref{G3})
represent both constraints and dynamic equations.
Below we solve all constraints and derive remaining independent
dynamic equations.

The first equation of motion, Eq. (\ref{E1}), is a constraint which allows to express
the longitudinal and transverse components of the pure gauge field $K_{00\delta}$ in terms of
corresponding components of $K_{\nu\nu\delta}$
\bea
&&K^l_{00\delta}=-\dfrac{1}{2}K^l_{\nu\nu\delta},  \\
&&K^{tr}_{00\delta}=-\dfrac{\rho}{\Delta+2 \rho} K^{tr}_{\nu\nu\delta}. \label{eqK00d}
\eea
To solve the Eqn. (\ref{E2}) it is convenient to
find first the longitudinal part of $K_{0\gamma\delta}$
\bea
&&\pro^\gamma K_{0\gamma\delta}=-\dfrac{\alpha \Delta}{\alpha \Delta+2 \gamma \rho}\pro^\beta K_{\delta 0\beta}.
\eea
With this one can resolve completely the constraint (\ref{E2})
\bea
K_{0\gamma\delta}&=&\dfrac{1}{(\alpha+\gamma)\Delta +2\gamma\rho}
      [-(\alpha+\dfrac{\alpha\gamma\Delta}{\alpha\Delta+2\gamma\rho})(\pro_\gamma\pro^\alpha K_{\delta 0\alpha}
                   -\pro_\delta\pro^\alpha K_{\gamma 0\alpha}) \nn \\
&-&((3\alpha-2\gamma)\rho-\gamma\Delta)(K_{\gamma 0\delta}-K_{\delta 0\gamma})]. \label{solto23}
\eea
The constraint (\ref{E3}) has a simple solution
\bea
&&K_0=\dfrac{1}{\Delta}\pro_0\pro^\alpha K^\nu_{~\nu \alpha}.\label{solto24}
\eea
This solution can be rewritten in terms of irreducible component fields
\bea
&&\Delta \TR=\dfrac{2\pro_0\pro^\alpha}{\Delta} A_\alpha . \label{eqRT}
\eea

The longitudinal projection of Eqn. (\ref{E4}) represents a constraint
which has the following solution
\bea
&& S^l_{\alpha}=-\dfrac{\gamma}{2(\alpha+\gamma)} \pro_\alpha \Delta \TS.
\eea
Using the above solutions one can check that
the transverse projection of the Eqn. (\ref{E4}) produces a
dynamic equation for the transverse field $K_{\nu\nu\delta}$
\bea
&& \Box K^{tr}_{\nu\nu\delta}+\rho K^{tr}_{\nu\nu\delta}+2 \rho K^{tr}_{00\delta}
           +\dfrac{2\gamma\rho}{\alpha\Delta}\pro_0 R^{tr}_\delta =0. \label{preeq1}
\eea
In the next section we will show that the original quadratic in $K_{mcd}$ Lagrangian
possesses additional local $U(1)$ and $\chi$ symmetries. One can check that the
field $K^{tr}_{\nu\nu\delta}$ is invariant under $\chi$-transformations and
represents two transverse modes corresponding to
$U(1)$ gauge boson. Notice, one should check the consistence of the Eq. (\ref{preeq1})
with all remaining  equations of motion which can constrain strongly the field dynamics of
$K^{tr}_{\nu\nu\delta}$.

Solving the Eqn. (\ref{E5}) is a little tedious but straightforward.
Notice, that we consider all equations
including $\rho$-dependence, i.e., we take into account
next linear order terms in $\rho$. One can verify that
in a case of constant space-time
background, $\rho=const$, there is no higher order corrections to the equations of motion.
The antisymmetric part of the Eqn. (\ref{E5})
implies two constraints on the irreducible field components
\bea
&&\Big (C_1-\dfrac{4C_2^2}{A}\Big )\pro^\alpha Q_\alpha +(2\gamma-\alpha)\pro_0\Delta\Delta \TS
 +4 \gamma \pro_0 \Delta \pro^\alpha S_\alpha=0, \label{solto26a} \\
 &&\Big [\dfrac{2C_2}{(\alpha+\gamma)\Delta +2 \gamma \rho}\big (-(\alpha+\gamma) \Delta +\dfrac{2\gamma^2 \rho}{\alpha}\big )
 +(-\alpha+\alpha\xi-\dfrac{2\gamma \Delta}{\Delta-6 \rho})\Delta \Big ]R^{tr}_\alpha= \nn \\
&&-4\alpha \rho\dfrac{\pro_0}{\Delta} A^{tr}_\alpha, \label{solto26b}
  \eea
where $C_1, C_2$ are operator quantities
\bea
C_1&=& 2 (\alpha+2 \gamma) \Delta +4 \alpha \pro_0 \pro_0 +2(4 \gamma-3\alpha) \rho, \nn \\
C_2&=& -\gamma \Delta+(3 \alpha-2 \gamma) \rho.
\eea
Notice, that the relationship (\ref{solto26b}) between the transverse components
$A^{tr}_\alpha$ and $R^{tr}_\alpha $
appears due to taking into account the presence of the curved space-time
with a non-zero curvature $\rho$.

Taking the divergence of the symmetrized part of the Eqn. (\ref{E5})
one can obtain another two constraints
\bea
&&2 \pro^\alpha R_\alpha=\pro_0 \pro^\alpha A_\alpha, \label{eqlong1} \\
&&\rho(1+\dfrac{\gamma}{3\alpha}) \Delta R^{tr}_\delta=0. \label{eqtr1}
\eea
Due to the imposed gauge condition (\ref{G3}) the Eqs. (\ref{eqlong1}) and (\ref{solto24}) imply
that $R^l_\alpha=A^l_\alpha=\TR=0$.
One should stress that the equation (\ref{eqtr1}) is non-trivial, and
it appears in linear order in curvature $\rho$.
An important consequence of the equations (\ref{eqtr1}) and (\ref{solto26b})
is that the field $A^{tr}_\alpha$ vanishes unless $\gamma = -3\alpha$.
This is the only case, $\gamma = -3\alpha$, when the field $A^{tr}_\alpha$
gains propagating modes.
Finally, taking into account the above solutions,
the symmetric traceless part of the Eqn. (\ref{E5})
leads to a relationship
between spin two fields $\TTR_{\beta\delta}$ and $\TTS_{\beta\delta}$
\bea
&&(\Delta +3 \rho) \TTR_{\beta\delta}-
 \dfrac{1}{2}(\epsilon_{i\delta}^{~~~\rho} \pro_0\pro^i \TTS_{\beta\rho}+
       (\beta \leftrightarrow \delta))=0. \label{RSrel}
\eea

Surprisingly, the general Lagrangian (\ref{lastL}) admits
dynamic equations for the contortion only with a
unique choice of the parameters, ($\beta=0, \gamma=-3, \alpha=1$),
($\alpha$ is unimportant overall number factor which can be set to one).

Let us rewrite the gauge conditions and solutions to the equations of motion
for the special case of chosen parameters $\alpha=1,\beta=0, \gamma=-3$ and small values of $\rho$.
Notice, the approximation $\rho\simeq 0$  corresponds to small deviation of the space-time
geometry from the flat limit which is enough for our purpose to analyze the dynamic properties
of the fields.
The gauge conditions (\ref{G3}) can be written in terms of independent fields
as follows
\bea
&&Q_\gamma^{tr}=0,  \\
&&S^{tr}_\alpha=\epsilon_{\alpha\gamma}^{~~~\delta} \dfrac{\pro^\gamma}{\Delta} A^{tr}_\delta, \label{eqStr} \\
&&S^l_\gamma=-\dfrac{3}{4}\pro_\gamma \TS, \\
&& R^l_\alpha =0.
\eea
The solution (\ref{solto23}) is simplified as follows
\bea
&&K_{0\gamma\delta}=-\dfrac{\Delta}{\Delta -6\rho}(\pro_\gamma R_\delta-
\pro_\delta R_\gamma)+3 \epsilon_{\gamma\delta}^{~~~\sigma}Q_\sigma.
\eea

Now we can write down the dynamic equation (\ref{preeq1}) for $A^{tr}_{\alpha}$
in a final form
\bea
(\Box +2 \rho) A^{tr}_\delta=0. \label{E4fin}
\eea
Since the fields $A^{tr}_\alpha$ and $S^{tr}_\alpha$ are related by the Eqn. (\ref{eqStr}),
the same equation holds for the field $S^{tr}_\alpha$.

The Eqns. (\ref{solto26a},\ref{solto26b}) can be rewritten in a simple form
\bea
&&2 \pro_0\pro_0 Q_\alpha+4(\Delta+3\rho) Q_\alpha +\pro_0\pro_\alpha \Delta \TS-
 \epsilon_{\alpha\beta}^{~~~\delta}( \pro_0 \pro^\beta K_{00\delta}+6\rho\pro^\beta R_\delta)=0, \label{E5.2} \\
 && R^{tr}_\delta = \dfrac{\pro_0}{3\Delta} A^{tr}_\delta.
\eea

Using the identity for the irreducible field $\TTS$
\bea
&& \TTS_{\alpha\beta}+\epsilon_{\alpha\gamma}^{~~~\delta}\epsilon_{\beta \nu}^{~~~\rho}
 \dfrac{\pro^\gamma \pro^\nu}{\Delta} \TTS_{\delta\rho} \equiv 0
 \eea
one can solve the last equation of motion, (\ref{E6}), which reproduces the eqn. (\ref{E4fin})
and leads to additional equations
\bea
&&(\Box +3\rho)(\pro^\alpha Q_\alpha-\dfrac{2}{3} \Delta \pro_0 \pro^\alpha S_\alpha)=0, \label{E6.1}\\
&&\rho \Box \TTS_{\alpha\beta}=0. \label{E6.7}
\eea
Notice, that there is no equation for spin two field $\TTS_{\alpha\beta}$ at zero order in $\rho$.
The non-trivial dynamics for the spin two field $\TTS_{\alpha\beta}$
appears only in linear order in $\rho$,
i.e., in the presence of curved space-time background.
The Eqns. (\ref{E5.2}, \ref{E6.1}) implies that the longitudinal components
of the vector fields $S_i, Q_i$ become propagating. Defining scalar
fields corresponding to the longitudinal components of $S_i, Q_i$
\bea
&&\varphi=\pro^\alpha Q_\alpha, \nn \\
&& \psi=-\dfrac{2}{3}\pro^\alpha S_\alpha,
\eea
one can easily derive the following dynamic
equations
\bea
&&(\Box+6\rho) \varphi +\Delta \sigma=j_1, \nn  \\
&& (\Box+6\rho) \psi+\pro_0 \sigma= j_2,  \nn \\
&& \sigma=\varphi+\pro_0 \psi,
\eea
where we introduce explicitly sources $j_1, j_2$ for the fields $\varphi, \psi$.
The linear dependent field $\sigma$ satisfies the standard hyperbolic equation
\bea
&&(\Box+3\rho) \sigma=\dfrac{1}{2} (j_1+\pro_0 j_2).
\eea
So that, our model has six dynamic degrees of freedom corresponding to
the irreducible field components
($\TTS_{\alpha\beta}, A^{tr}_\alpha,\varphi, \psi$) with spins
($2;1;0;0$) respectively. One can treat the field $S^{tr}_\alpha$
as an independent one, since the fields $A^{tr}_\alpha, S^{tr}_\alpha$ are related by Eq.
(\ref{eqStr}). Then, the scalar $\psi=-2/3 \pro^\alpha S_\alpha$
represents a longitudinal component of the vector $S_\alpha$.

Notice, that the kinetic terms for the fields $A^{tr}_\alpha$ and $\varphi$
are positively defined, whereas the terms for the fields $\TTS_{\alpha,\beta}$ and $\psi$
have negative contribution to the classical Hamiltonian.
However, one should stress that the action might have still
positive definiteness since the theory has highly non-linear structure.
For instance, the kinetic term for the field $\TTS_{\alpha,\beta}$
contains the curvature $\rho$ as a multiplier, so that, it
represents actually the interaction term between the vielbein and contortion.

\section{Covariant quantization and the effective Lagrangian}

One-loop effective action with a constant curvature space-time
background and quantum torsion has been calculated recently
in the model with Yang-Mills type Lagrangian quadratic
in Riemann-Cartan curvature\cite{pak}.
In the previous section we have demonstrated that the
model given by the Lagrangian (\ref{lastL}) $(\alpha=1, \beta=0, \gamma=-3)$
has dynamic contortion.
In this section we perform covariant quantization of the model
which is more suitable for practical calculation of a quantum
effective action.

We apply the functional integral formalism to derive the
quantum effective Lagrangian in one-loop approximation.
In background field formalism one starts with splitting
the general gauge connection $A_{m cd}$
into background (classical) and quantum parts
\bea
&A_{m cd}= A^{({\it cl})}_{m cd}+A^{({\it q})}_{m cd}. \label{splitting}
\eea
We identify the classical field $A^{{\it (cl)}}_{m cd}$
with the Levi-Civita connection $\varphi_{mcd}(e)$
corresponding to the Riemannian space-time geometry
and the quantum field $A^{{\it (q)}}_{m cd}$
with contortion $K_{m cd}$.

Let us define two types of Lorentz
gauge transformations consistent with the
original Lorentz gauge transformation (\ref{eq:deltaA}) and the decomposition
(\ref{splitting}): \newline
(I) the classical, or background, gauge transformation
\bea
&&\delta e_a^m = \Lambda_a^{~b} e_b^m, \nn \\
&& \delta {\boldsymbol \varphi}_m(e) = -\pro_m {\bf \Lambda}-
                 [{\boldsymbol \varphi}_m,{\bf \Lambda}], \nn\\
&& \delta {\bf K}_m = -[{\bf K}_m,{\bf \Lambda}],\label{eqI}
\eea
(II) the quantum gauge transformation
\bea
&&\delta e_a^m =
\delta {\boldsymbol \varphi}_m(e)=0, \nn \\
&& \delta {\bf K}_m= - \hat D_m {\bf \Lambda}-[{\bf K}_m,{\bf \Lambda}],\label{eqII}
\eea
where ${\boldsymbol \varphi}_m \equiv \varphi_{m cd} \Omega^{cd}$, and
the restricted covariant derivative $\hat D_m$ is defined
by means of the Levi-Civita connection only
\bea
&&\hat D_m {\bf\Lambda}=\pro_m {\bf\Lambda}+[{\boldsymbol \varphi}_m,{\bf \Lambda}].
\eea
Notice that the restricted derivative $\hat D_m$ is covariant under
the classical Lorentz gauge transformation.
An interesting feature of the Lagrangian (\ref{lastL}) is its invariance
under above two types of gauge transformations. In general, this invariance
can be broken by interaction of contortion with matter fields.

In one-loop approximation it is sufficient to keep
only quadratic in contortion terms in the Lagrangian.
After integration by part and neglecting surface terms
the quadratic Lagrangian (\ref{lastL}) can be reduced to the form
\bea
{\cal L}_0^{(2)}&=&\hD_a K_{bcd}(\hD^a K^{bcd}-\hD^b K^{acd})
+4 \hD_a K_{bcd}\hD^c K^{dab}+ \nn \\
&&+3 \hD_a K_{bcd}(\hD^a K^{cdb}-\hD^c K^{adb}
-\hD^b K^{cda}+\hD^c K^{bda}). \label{lagrini}
\eea
The Riemann curvature is supposed to be covariant constant, i.e.,
$\hat D_a \hR_{bcde}=0$.

An interesting property of the quadratic Lagrangian (\ref{lagrini})
is the presence of an additional local $U(1)$ symmetry
\bea
&& \delta_{U(1)} K_{bcd} = \dfrac{1}{3} (\eta_{bc}\hD_d\lambda-\eta_{bd}\hD_c \lambda),\nn \\
&& \delta_{U(1)} K_{d} = \hD_d \lambda.  \label{U1}
\eea
As it was shown in the previous section using the Lagrange formalism,
the vector field $K_d$ contains two transverse dynamic modes,
$A^{tr}_\delta$. Notice, that a quadratic part of the Yang-Mills type Lagrangian
\bea
&& {\cal L}_{YM}=-\dfrac{1}{4} R_{abcd}^2 \label{LYM}
\eea
does not have such a local $U(1)$ symmetry.

It is convenient to decompose the contortion into
irreducible parts
\bea
&& K_{bcd}=Q_{bcd}+\dfrac{1}{3} (\eta_{bc} K_d-\eta_{bd} K_c)
+ \dfrac{1}{6} \epsilon_{bcde} S^e, \nn \\
&& Q^c_{~cd}=0,  \nn \\
&& \epsilon^{abcd} Q_{bcd}=0.
\eea
Another unexpected feature of the quadratic Lagrangian ${\cal L}^{(2)}$,
(\ref{lagrini}), is that it admits another local symmetry.
The symmetry is provided by the following
transformations with a new constrained parameter
$\chi_{bc}$
\bea
&& \delta_\chi Q_{bcd}= \hD_c \chi_{db}-\hD_d \chi_{cb}, \nn \\
&& \delta_\chi K_d=0, \nn \\
&& \delta_\chi S^a=0, \nn \\
&& \chi_{bc}=\chi_{cb},~~~\chi^c_{~c}=0, ~~~\hD_c\chi_{cd}=0. \label{eqchi}
\eea
The presence of the additional local gauge symmetries
implies that the model is degenerate and has additional constraints.
The field $Q_{bcd}$ has sixteen field components in general.
After subtracting six pure gauge degrees
of freedom due to Lorentz gauge symmetry and five degrees
due to $\chi$-symmetry one has exactly five physical
degrees of freedom for the spin two field.
The fact that we have only one physical spin two field
is unexpected, and it does not occur in the gauge gravity model with Yang-Mills type
Lagrangian ${\cal L}_{YM}$, (\ref{LYM}), where the contortion
$Q_{bcd}$ contains a pair of spin two fields.

To perform one-loop quantization one has to fix the
gauges corresponding to the local Lorentz, $\lambda$ and $\chi$
symmetries.
The simplest gauge fixing function
we have chosen is the following
\bea
&&F_{1cd}=\hD^b Q_{bcd}.
\eea
One has a simple Lorentz gauge transformation rule for the function
\bea
&& \delta F_{1cd} = -\dfrac{2}{3} ( \hD\hD+\dfrac{\hR}{6}) \Lambda_{cd}.
\eea
With this one can write down
the corresponding gauge fixing term and Faddeev-Popov ghost
Lagrangian
\bea
&& {\cal L}^{(1)}_{gf}=-\dfrac{1}{2 \xi_1} (\hD^b Q_{bcd})^2,  \nn \\
&& {\cal L}^{(1)}_{FP}=\bar c_1^{cd} (\hD\hD+\dfrac{\hR}{6}) c_{1cd},
\eea
where $\bar c_1^{\,cd}, c_{1cd}$ are ghost fields.
For simplicity we choose the gauge parameter $\xi_1=1$.
Notice, that the gauge function $F_{1cd}$ is invariant under
$U(1)$ and $\chi$-transformations.
To fix the gauge for the local $U(1)$ symmetry
one has to introduce a second gauge fixing function
which can be chosen as
\bea
F_2=\hD^b K_b.
\eea
The corresponding gauge fixing term and Faddeev-Popov Lagrangian
have the following form
\bea
&& {\cal L}^{(2)}_{gf} = -\dfrac{1}{\xi_2} (\hD^b K_b)^2, \nn \\
&& {\cal L}^{(2)}_{FP} = \bar c_2 \hD\hD c_2.
\eea
We choose a Feynman gauge ($\xi_2=1$) for simplicity.
Finally, to fix the gauge for the $\chi$-transformations one
can choose a constrained gauge fixing function
\bea
&& F_{3bd}=\dfrac{1}{2}(\hD^a Q_{bad}+\hD^a Q_{dab}), \nn \\
&& \eta^{bd}F_{3bd}=0, \nn \\
&& \hD^b F_{3bd}=\dfrac{1}{2} \hD^a\hD^b Q_{bad} \simeq 0,
\eea
where the last equality takes place on the hypersurface $\hD^b Q_{bcd}=0$
in the configuration space of functions $\{Q_{bcd}\}$.
One can easily find the corresponding gauge fixing and ghost terms
\bea
&& {\cal L}^{(3)}_{gf}=-\dfrac{1}{2\xi_3} F_{3bd}^2, \nn \\
&& {\cal L}^{(3)}_{FP}=\bar \psi^{cd} (\hD\hD-\dfrac{\hR}{3}) \psi_{cd}. \label{tachyon}
\eea
We set condition $\xi_3=\dfrac{1}{2}$ which corresponds to a
symmetric gauge. In calculation of the one-loop effective action
after functional integration over the ghost fields $\psi_{cd}$
the last equation (\ref{tachyon})
leads to a functional determinant of the operator
$\hD\hD-\hR/3$ which has an additional pole (for a
constant positive curvature $\hR$). This implies the presence
of the tachyon mode which has the same origin as the known
Savvidy-Nielsen unstable mode
in quantum chromodynamics\cite{savv,niel}.

The final expression for a total one-loop
effective Lagrangian is given by the sum
of all gauge fixing and ghost terms
\bea
&& {\cal L}^{(2)}_{eff} ={\cal L}^{(2)}_{0}+\sum_{i=1,2,3}({\cal L}_{gf}^{(i)}
                  +{\cal L}_{FP}^{(i)}).\label{eq:L2}
\eea
The expression for the effective Lagrangian is ready for
calculation of the one-loop effective action. The contributions
of ghosts  are given by scalar functional
determinants which can be easily calculated in analytic form\cite{pak}.
Unfortunately, the calculation of the functional determinants
obtained after integration over contortion is more complicate due to
the tensorial structure of the corresponding propagator.

\section{Discussion}

In our previous paper\cite{pak} we have proposed a mechanism
of dynamical generation of Einstein gravity through the quantum corrections
of torsion in the Yang-Mills type Lorentz gauge gravity model.
The main ingredient of such a mechanism is the formation of
torsion vacuum condensate which we assumed to be
covariant constant
\bea
&&<\tilde R_{abcd}>= \difrac{1}{2} M^2(\eta_{ac} \eta_{bd}-\eta_{ad}\eta_{bc}),
\label{assumpn}
\eea
where $M^2$ is a mass scale characterizing the
torsion condensate. We suppose that the model proposed in the
present paper admits the generation of the Einstein-Hilbert
and cosmological terms as well.

In conclusion, we propose a simple $R^2$ model of Lorentz gauge
quantum gravity with torsion.
Our model has a number of advantages to compare
with Yang-Mills type Lorentz gauge gravity.
In the absence of classical torsion the model reduces to a pure
topological gravity, i.e., one has a topological phase
where the metric is not specified a priori. The metric
can obtain its dynamical content after
dynamical symmetry breaking in the phase of the
effective gravity which includes the Einstein gravity as its part.
An unexpected feature of our model is that the contortion has the same number
of dynamic degrees of freedom as the number of physical components of the
metric tensor in Einstein gravity. The difference is that in Einstein gravity the metric has
only two propagating spin 2 modes, other polarization modes corresponding to spin
states $(1;0;0)$ are not dynamical. Whereas in our model
the contortion has six propagating modes with spin polarizations
$(2;1;0;0)$, that means the torsion might have a dynamical mass
like a gluon in quantum chromodynamics.
This fact can be considered as an additional hint that the contortion
could be a quantum counter-part to the classical graviton.

We have analyzed the dynamic structure of our model
considering linearized equations of motion
in a constant curvature space-time.
It should be noted, that our results can reflect the principal dynamic properties of
the full non-linear theory since the Levi-Civita connection
can be treated as a background part of the Lorentz gauge connection.
A more detailed analysis of the non-linear structure
of our model and interaction with matter will be considered in a separate
forthcoming paper.

\section*{Acknowledgments}

One of authors (DGP) thanks Prof. K.-I. Kondo for interesting
discussions. The work is supported in part by the ABRL Program of
Korea Science and Engineering Foundation (R14-2003-012-01002-0)
and by Brain Pool Program (032S-1-8).

\end{document}